\documentclass[jkps,preprint,fleqn,showpacs,showkeys]{revtex4}
\usepackage{graphicx}
\usepackage{amssymb}
\usepackage{amsmath}
\usepackage{bm}
\begin{document}
\setcounter{page}{0}
\title[]{Structural Properties of Networks Grown via an Achlioptas Process}
\author{Woo Seong \surname{Jo}}
\affiliation{Department of Physics, Sungkyunkwan University, Suwon 440-746, Korea}
\author{Su Do \surname{Yi}}
\affiliation{Department of Physics, Pukyong National University, Busan 608-737, Korea}
\author{Beom Jun \surname{Kim}}
\email{beomjun@skku.edu}
\affiliation{Department of Physics, Sungkyunkwan University, Suwon 440-746, Korea}
\author{Seung-Woo \surname{Son}}
\email{sonswoo@hanyang.ac.kr}
\affiliation{Department of Applied Physics, Hanyang University, Ansan 426-791, Korea}


\begin{abstract}
After the Achlioptas process (AP), which yields the so-called explosive percolation,
was introduced, the number of papers on percolation phenomena has been
literally exploding. Most of the existing studies, however, have focused only on
the nature of phase transitions, not paying proper attention to
the structural properties of the resulting networks, which compose
the main theme of the present paper. We compare the resulting network structure of the AP
with random networks and find, through observations of the distributions of the
shortest-path length and the betweenness centrality in the giant cluster, that the AP makes the
network less clustered and more fragile. Such structural characteristics are more directly seen by using snapshots of the network structures and are  explained by the fact that the AP suppresses the formation of large clusters more strongly
than the random process does. These structural differences between the two processes are shown to be
less noticeable in growing networks than in static ones.
\end{abstract}


\maketitle

\section{INTRODUCTION}
After the explosion of interest in complex network research that started in the
late 20th century, various complex network models such as the Watts-Strogatz
small-world network~\cite{Watts1998} and the Barab\'asi-Albert scale-free
network~\cite{Barabasi1999} have been intensively investigated in order to properly
describe the structures of connections in technological, social, and biological systems,
with a focus on the connection topology in them~\cite{AlbertRMP, Dorogovtsev2002, Newman2003}.
The simplest network model is the Erd\"os-R\'enyi (ER) random
network~\cite{Erdos1959}, where $N$ isolated nodes are connected randomly with
probability $p$~\cite{note0}. The ER random network is one of the important network
models in statistical physics and has been widely adopted in the analyses of the
properties of complex networks and the study of percolation theory.

Because of the simplicity and the mathematical tractability of the ER random
network, its structural properties have been well studied analytically and
numerically~\cite{Newman_text}. Because each node has equal probability to
have a link in the ER network, the biggest cluster comparable to the system size
$N$ emerges when the average link density (the fraction of the number of
links compared to $N$) approaches $1/2$ in the limit of $N \longrightarrow \infty$.
The network is said to be percolated when such a giant cluster is formed, and
the order parameter, defined as a number of nodes in the largest cluster compared to $N$,
starts to increase after the link density passes a critical value from below.
The growth behavior of the giant cluster follows a power-law with a
critical exponent $\beta = 1$ in ER random networks~\cite{Bollobas1985,
Dorogovtsev2008}, hence undergoing continuous phase transition.  The values of
the critical link density and the exponent $\beta$ have been the main research
interest, igniting active discussion on the universality class of the
continuous/discontinuous percolation transition in various model
systems~\cite{Dorogovtsev2008, Achlioptas2009, Cho2009, daCosta2010,
Grassberger2011, Lee2011, Cho2011, Riordan2011, SDYi2013, Cho2013}.

Recently, percolation phenomena have been investigated in various complex
network structures, revealing various transition natures that depend on how
the networks are built~\cite{Achlioptas2009, Cho2009, daCosta2010, Grassberger2011,
Lee2011, Cho2011, Riordan2011, SDYi2013, Cho2013}. In particular,
the celebrated Achlioptas process (AP)~\cite{Achlioptas2009} has drawn
enormous attention due to the existence of a very sharp transition,
as the name `explosive percolation' suggests. The explosive nature
of the transition occurs because the AP suppresses the formation of
large clusters by avoiding the linkage between larger clusters~\cite{note2}.
After an initial debate about the true nature of the explosive transition
(discontinuous or continuous), the scientific community has approached an
agreement that the explosive percolation transitions in the original AP and
its variants are, indeed, continuous~\cite{daCosta2010,
Grassberger2011, Lee2011, Riordan2011}.

Different from the static networks in which the number of nodes is fixed,
many networks in the real world are changing in time, and their numbers
of nodes and links often grow as time evolves.
In the Callaway model~\cite{Callaway2001}, which appears to be the simplest one
among growing network models, a node is added every time step, and a link
is added with the probability $\delta$ at random.
Many studies on various growing networks also
focus on the emergence of the giant cluster and on network properties like
the degree distribution and the clustering
coefficient~\cite{Callaway2001,Krapivsky2000,Zalanyi2003}.

Percolation in growing networks with a random merging process
is known to exhibit an infinite-order phase transition~\cite{Callaway2001},
in which the order parameter around the critical point changes very smoothly
with any order of derivative being continuous, analogous to
the Berezinskii-Kosterlitz-Thouless transition in the two-dimensional
$XY$ model~\cite{Dorogovtsev2008}. The percolation transition in growing scale-free networks with preferential
attachment has also been reported to show an infinite-order phase transition due to the aging
effect of early nodes~\cite{Dorogovtsev2001}. Recently,
percolation in growing network under an AP was studied, and
the second-order phase transition was found to exist. This can be
understood from the competition between the aging effect from growing
and the abrupt phase transition due to the AP without growing~\cite{SDYi2013}.

While many studies on explosive percolation have aimed to reveal the
nature of the phase transition, we instead investigate in the present paper
the structural properties of networks grown via an AP
in comparison to those of randomly-grown networks.
We also consider both static and growing networks. In this paper, we look into
several network properties such as the degree distribution, the clustering coefficient,
the shortest-path length distribution, and the betweenness centrality characteristics.
Furthermore, we test the robustness of the networks against errors and attacks.

\section{Method}

In this study, we consider four different types of networks with two criteria:
whether the network is static or growing, and whether the
link attachment rule is a random process (RP) or an Achlioptas process (AP). We
distinguish all four possible combinations with the terms static RP (SRP),
growing RP (GRP), static AP (SAP), and growing AP networks (GAP). The SRP networks are
fully identical to the standard ER random networks~\cite{Erdos1959}, and the GRP
networks exactly correspond to the Callaway model~\cite{Callaway2001}.
Both for SAP and GAP networks, we use the AP proposed by da
Costa et al. for simplicity. Consequently, the SAP and the GAP
networks in this work are identical to the networks in Ref.~\cite{daCosta2010}
and in Ref.~\cite{SDYi2013}, respectively.

For static network construction, we start from $N$ isolated nodes,
and at each step, two nodes chosen by using a specific link attachment rule, RP or AP,
are connected. In this study, we fix the total number of links $L$ as the system size $N$
(i.e., the number of nodes), which corresponds to the link density $\delta=L/N=1$~\cite{note1}.
To implement RP, we randomly chose two nodes from $N$ nodes. If a link connecting the pair already exists,
another two nodes are randomly chosen and then connected.
For the AP, on the other hand, two nodes are chosen
randomly, and the one that belongs to the smaller cluster is picked.
The same procedure is repeated once more to select the other node to connect.
The two selected nodes are then connected if they are an unoccupied pair~\cite{daCosta2010}.
The above procedure for static network construction needs to be altered for
growing networks: We start from a single node, and at each time step, an isolated node is added,
and either the RP or the AP is applied. The procedure is repeated until the network size reaches
the target $N$~\cite{SDYi2013}.

To describe the details of GAP: Starting from a single node, at each time step, we add a new node, and a link is added, which corresponds to the link density $\delta=1$. Instead of randomly adding a link to a pair of nodes like the RP, we choose a link according to the AP. First, we choose two nodes uniformly at random, pick the node that belongs to the smaller cluster, and repeat the same one more time. Then, the two selected nodes are connected~\cite{SDYi2013}.

Our aim here is to see how the network structures differ for the four different
cases, SRP, SAP, GRP, and GAP. We compare the structural properties by measuring
various quantities such as the degree distribution~\cite{AlbertRMP}, the clustering coefficient (CC)~\cite{Serrano2006, CYLee2006},
the shortest-path length (SPL) distribution~\cite{Freeman1977, SWSon2009}, and the betweenness centrality (BC) distribution~\cite{Freeman1977, Brandes2001, SWSon2004}. In order to check the robustness of the different networks, we
also test the tolerance against errors and attacks.

The robustness of the network is characterized by how much the network is tolerant
to damage, which is estimated by using the size of the remaining network after node (or
link) removal. It has been studied for various network models~\cite{AlbertRMP,
BJ2002, Callaway2000, Albert_attack, Cohen2001} and real-world networks like the World Wide Web~\cite{Albert_attack} and scientific collaboration networks~\cite{BJ2002}.
The robustnesses of scale-free networks~\cite{Cohen2001} and random
networks~\cite{Callaway2000} were analyzed theoretically from the viewpoint of
percolation. Network topology has turned out to be important for understanding the robustness. For example, the scale-free network is more robust
than the random network against the random removal of nodes or links (called `error').
In contrast, the former becomes particularly fragile against
preferential node removal (called `attack'), meaning that
removal of hub nodes can cause the network to lose a significant portion of
the connected components~\cite{Albert_attack}.
Later, the concept of a bicomponent,
in which each and every pair of nodes has at least two independent
connecting paths, was suggested to play
an important role as a robustness measure~\cite{Newman2008}.
Networks with larger bicomponents are robust by virtue of the existence
of backup pathways. We point out that an investigation of the network's robustness can help
in understanding the structures of the networks.

\section{Results}

\begin{figure*}[t]
\includegraphics[width=\textwidth]{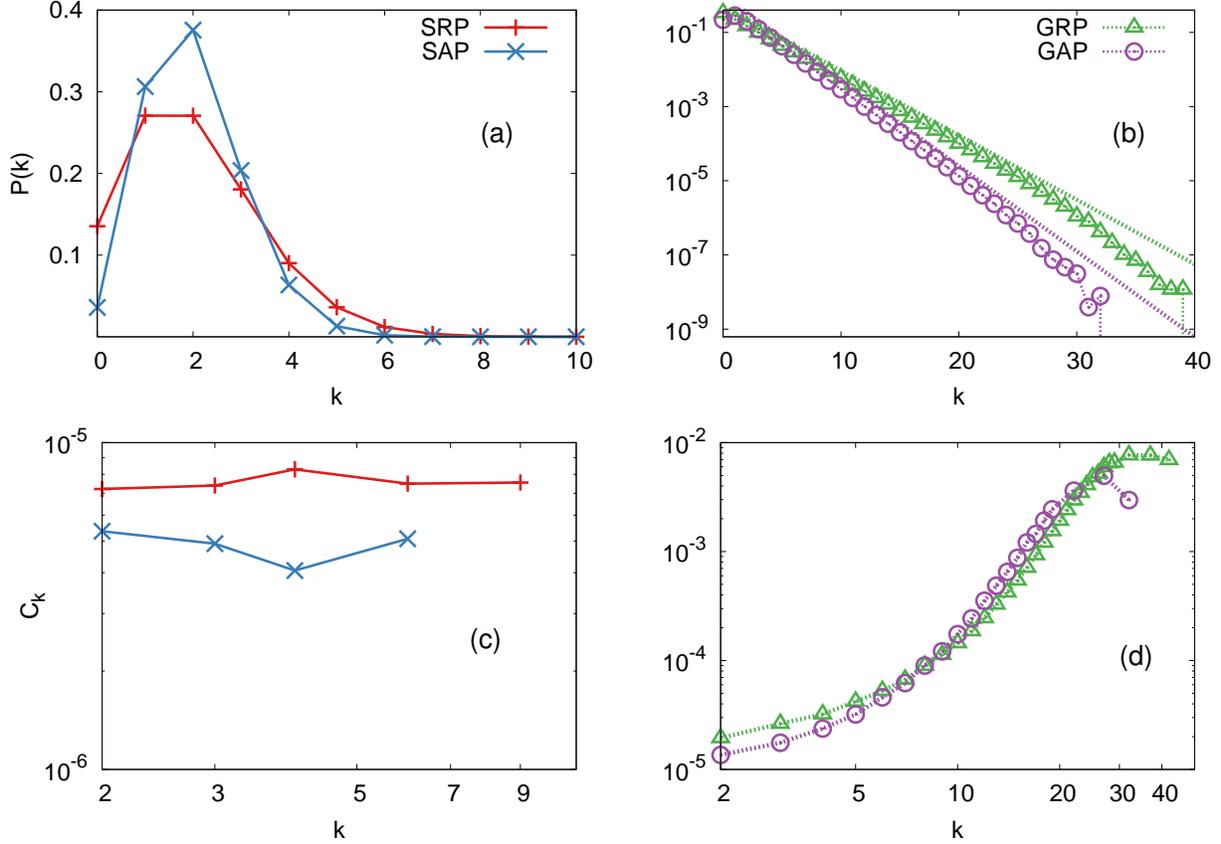}
\caption{Degree distribution $P(k)$ of (a) static networks (SRP and SAP) and
(b) growing networks (GRP and GAP). Local clustering coefficient (CC) $C_k$ versus the degree $k$
for  (c) static networks (SRP and SAP) and (d) growing networks (GRP and GAP). The system size
$N=256\,000$, and all observables are from averages over $1\,000$ network realizations. The
sizes of the error bars (not shown) are smaller than the symbol sizes. } \label{fig:dist&ck} \end{figure*}

First, we investigate the degree distribution, the most basic structural
property, of the four networks. The degree distribution
of the SRP network, i.e., the ER random network, is well known to follow the Poisson
distribution~\cite{Newman_text}, and that of the GRP network is well known to follow an exponential
one~\cite{Callaway2001}, as confirmed in Figs.~\ref{fig:dist&ck}(a) and (b),
respectively, for $\langle k \rangle =2$~\cite{note1}.
Both the SAP and the GAP networks also appear to follow the Poisson
and the exponential distributions, respectively [Figs.~\ref{fig:dist&ck}(a) and (b)].
Suppression of the formation of big clusters in the AP will also surely suppress the
appearance of high-degree nodes effectively, which is
reflected in the large-degree part of the distributions, as seen in
Figs.~\ref{fig:dist&ck}(a) and (b). Likewise,  small clusters,
like unconnected single nodes, have more chance to form a cluster, lowering
the frequency of nodes of very small degrees in comparison to the RP; this can
also be seen in the small-degree part of the distributions
in Figs.~\ref{fig:dist&ck}(a) and (b).
Consequently, the degree distributions of the AP networks
become narrower than those for the RP networks.
It is interesting to note that although RP and AP yield
strikingly different natures in the percolation phase transitions, the
supercritical degree distributions are not much different from each other.

Next, we examine how clustered each network is by measuring the CCs. The global
CCs of the AP networks are found to be lower than those of the RP networks for both static
and growing networks. In Figs.~\ref{fig:dist&ck}(c) and (d), we plot the local CC
as a function of the degree $k$, i.e., $C_k = \frac{2 \times N_\triangle}{k(k-1)}$, where $N_\triangle$ is the number of triangles connected to the vertex. In static networks [see
Fig.~\ref{fig:dist&ck}(c)], the SAP is shown to have consistently lower $C_k$ than the SRP.
Note that the $C_{k}$ for SRP networks is well described by the expected
result, $k / (N-1) \sim 7.8 \times 10^{-6}$~\cite{Newman_text}. For growing
networks, the GRP and the GAP appear to have qualitatively the same shapes for $C_k$.

\begin{figure*}[t]
\includegraphics[width=\textwidth]{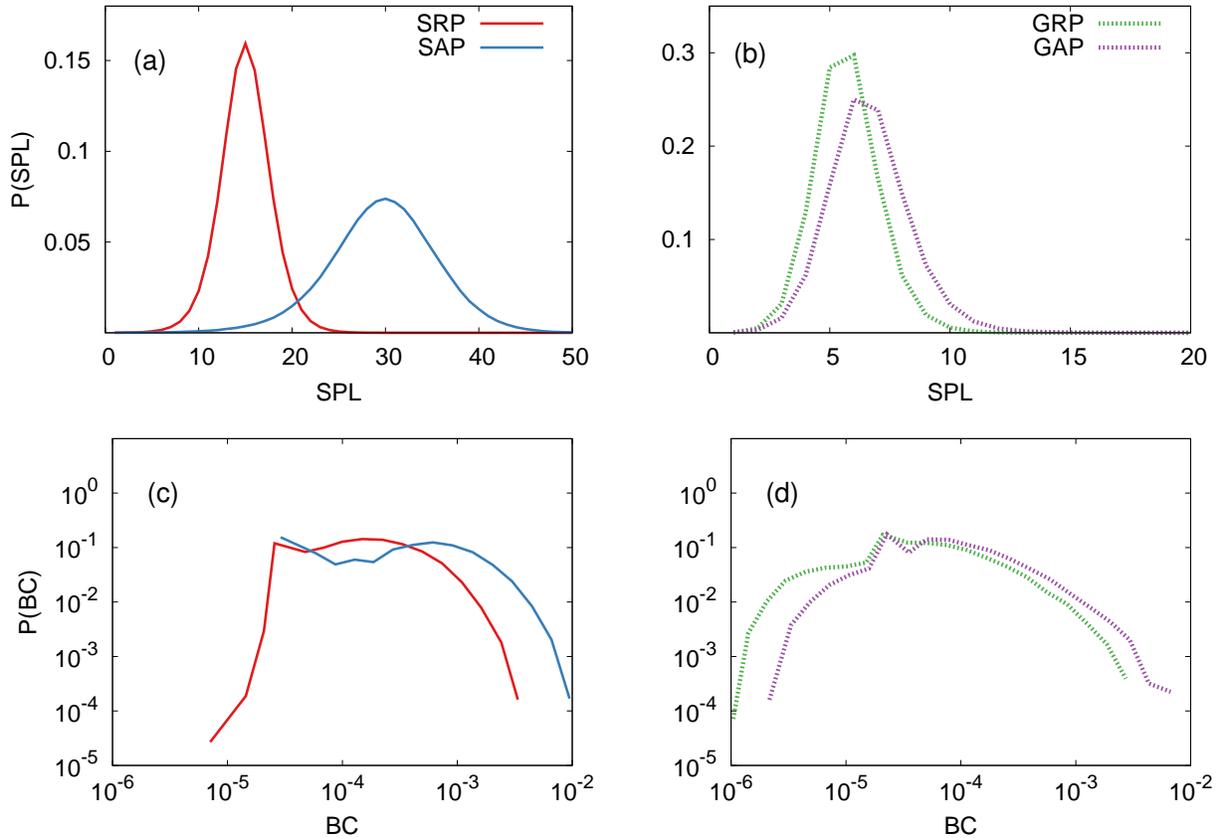}
\caption{Shortest-path length (SPL) distribution for (a) static networks and
(b) growing networks. Betweenness centrality (BC) distribution for (c)
static networks and (d) growing networks. $N=64\,000$ was used for calculations.
}
\label{fig:Path}
\end{figure*}

In order to understand the structural differences of the four networks
from another perspective, we measure the distributions of the SPL, the minimum hopping distance between connected pairs, and the BC, the average amount of traffic passing through the specific node
computed for the nodes in the largest connected component (calculations of
the two quantities is meaningless for disconnected networks). From  now on,
we use a smaller network size $N = 64\,000$ because of the computation complexity
in larger networks. Even though the AP does not change the
degree distribution and the local CCs significantly, it obviously alters the
distributions of the SPL as shown in Figs.~\ref{fig:Path}(a) and (b). When the AP
is applied, the characteristic path lengths, the average of the SPLs, is larger
than that of the RP networks. The increase in the characteristic path length
is more noticeable for static networks, which means the SAP network is more chain-like
than the SRP network.

Figures~\ref{fig:Path}(c) and (d) show that the largest BCs of the AP networks are
larger than those of the RP networks both for (c) static and (d) growing networks.
This is particularly interesting because RP networks have thicker tails in
the degree distributions (Fig.~\ref{fig:dist&ck}), in sharp contrast to the thinner
BC distributions for the RP (Fig.~\ref{fig:Path}). For a network with weak community structures, the higher the node
degrees are, the larger are the BCs, in general. In contrast, some nodes in networks with well-structured communities
can have larger BC, and smaller degrees at the same time. Such nodes
are identified as inter-community nodes and connect different communities.
In the AP, due to its suppression of the formation of
large clusters, separate clusters of similar sizes can coexist.
While the network is growing, small clusters merge to
form a large one, and some node can become an inter-community node by making a connection
between clusters.  The existence of such nodes is exhibited in
the tail part of the BC distribution (see Figs.~\ref{fig:Path}(c) and
(d)). The difference in the BC distributions between AP and RP networks is also more
noticeable in static cases than in growing ones.

\begin{figure*}[t]
\includegraphics[width=0.48\textwidth]{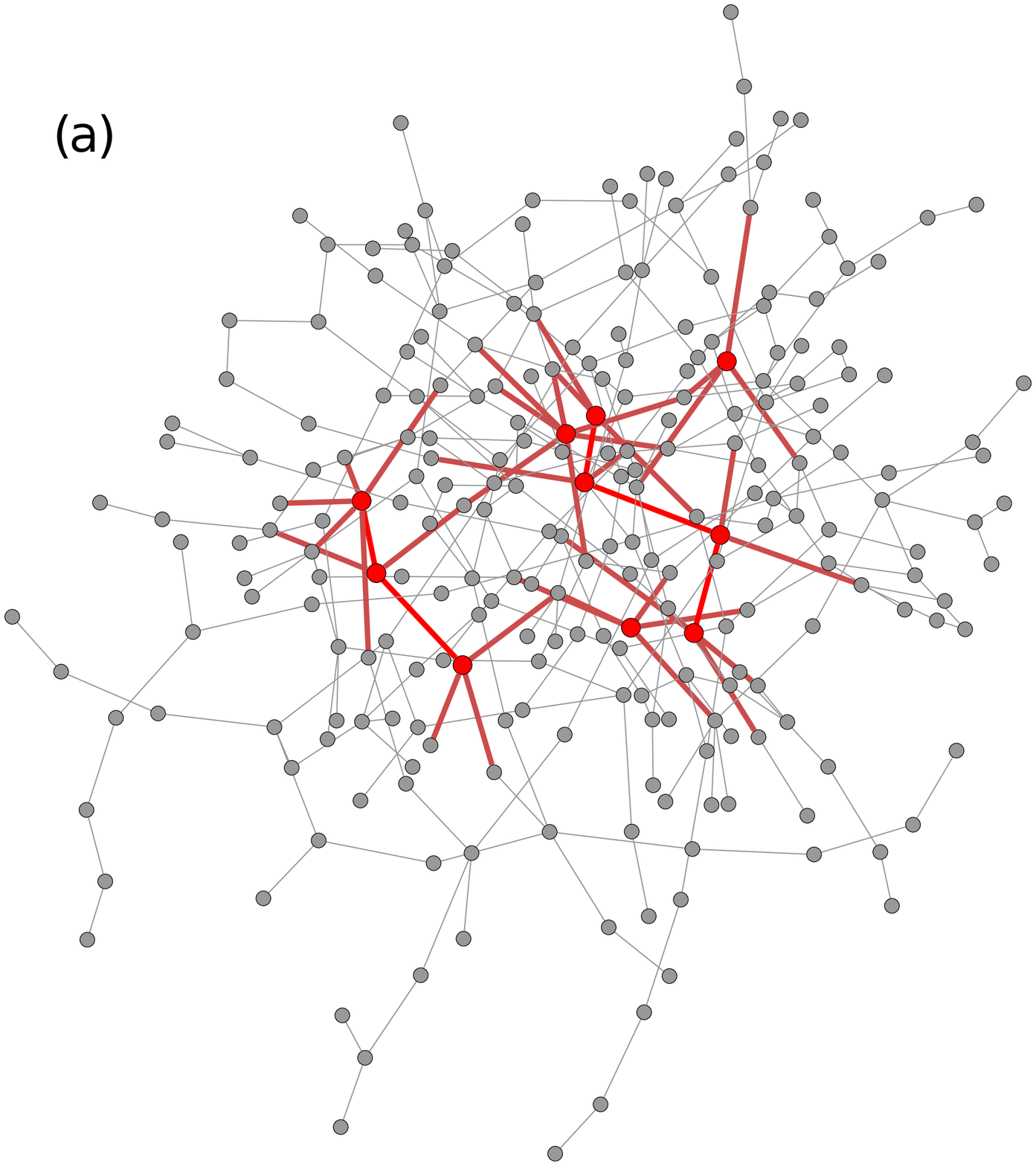}
\includegraphics[width=0.48\textwidth]{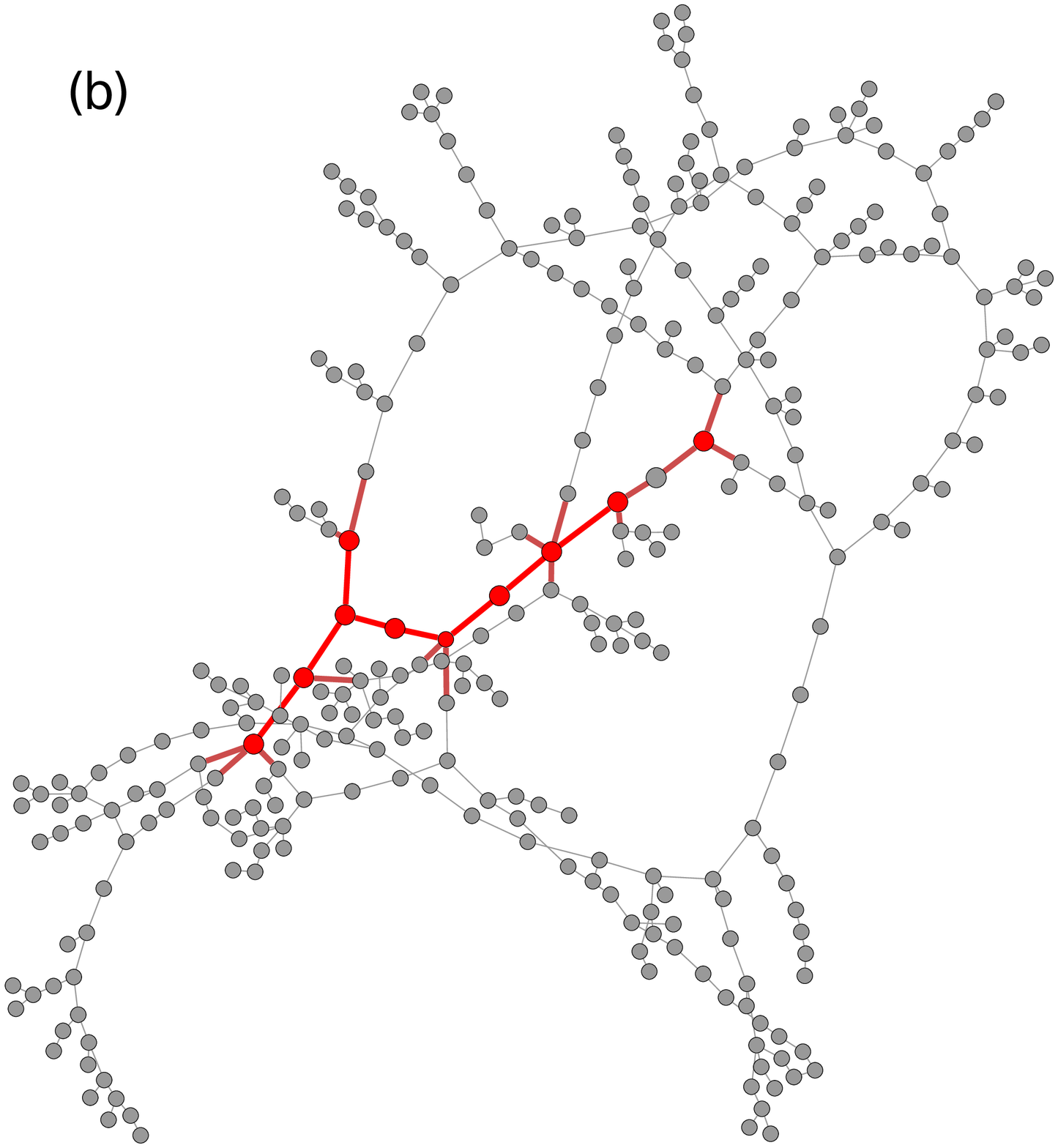}
\caption{Snapshots of (a) a SRP network and (b) a SAP network. For better visibility,
networks with only a few hundreds of nodes are displayed. The 10 highest BC nodes and their links are marked red.
}
\label{fig:snapshots}
\end{figure*}

These changes in the SPL and the BC distributions under the AP clearly imply that the
structures of AP networks are different from those of RP networks even though one
hardly notices the changes in $P(k)$ and $C_k$. As an example, Fig.~\ref{fig:snapshots}
shows snapshots of the giant components in the SRP and the SAP networks. One can
see that many nodes are connected along the chain-like structure in an AP network.
On the contrary, the giant cluster in a RP network is more interwoven with more tangled
links, which provide shortcuts while calculating the SPL and the BC. Here, the 10
highest BC nodes and their links are marked in red. Note that even though the growing networks also show little difference between the GRP and the GAP, distinguishing the structural difference between them by looking at the snapshots is hard because the growing effect interweaves a quite-strong, single, giant component.

\begin{figure*}[t]
\includegraphics[width=\textwidth]{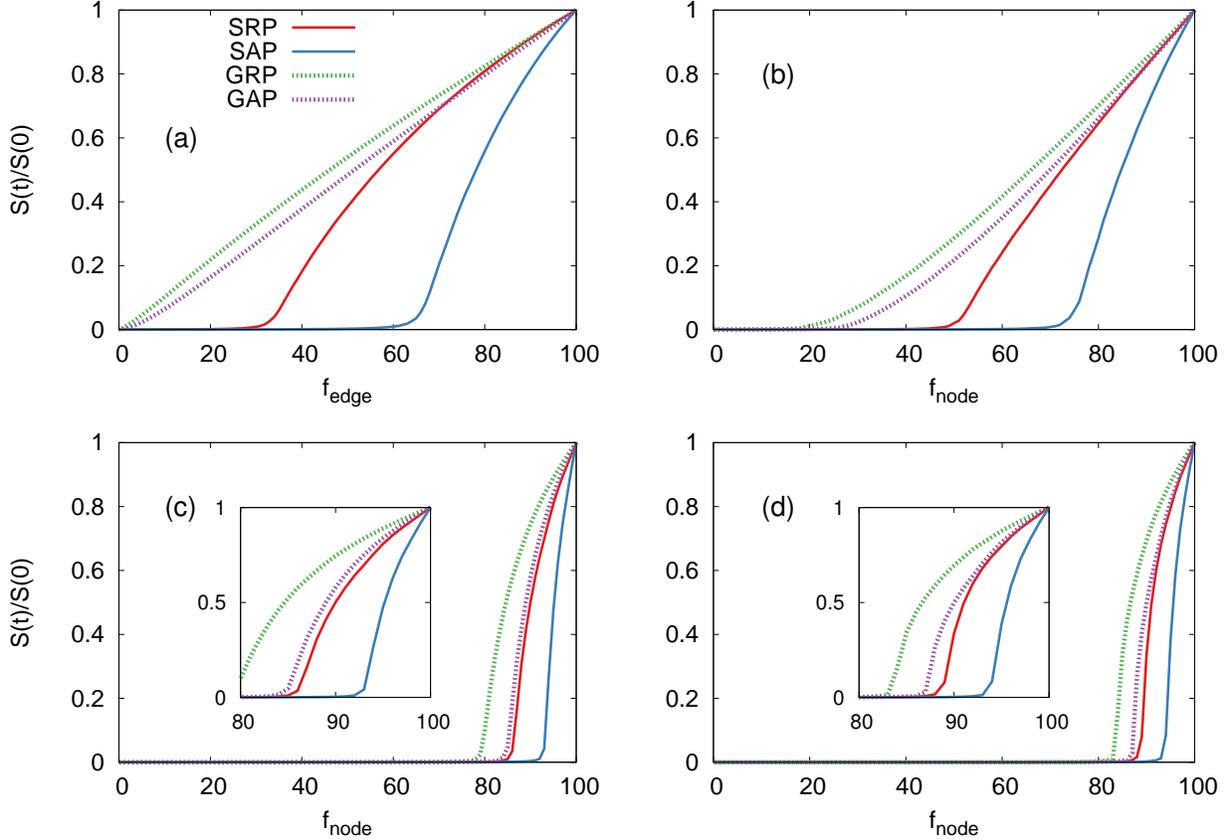}
\caption{Robustness against errors and attacks. The fraction of the largest
cluster, $S(t)/S(0)$, is observed. Random removal of (a) links and (b) nodes as
errors ($N=64\,000$). $f_{edge}$ and $f_{node}$ denote the proportions of remaining
edges and nodes at time $t$, respectively. Targeted removal of large-degree nodes by using (c)
the initial degree without recalculating and (d) the remaining degree by recalculating
($N=64\,000$).}
\label{fig:Attacks}
\end{figure*}

Finally, we test the vulnerability of the RP and the AP networks by performing random link (or node) deletion, and the targeted node was removed based on its degree.
In network research on robustness, the former disturbance of a given
network is usually called `errors' and the latter `attacks.' After removing a certain
fraction of nodes in the networks, we measure the ratio between the size of the largest
cluster $S(t)$ at the $t$-th step of deletions and its initial size $S(0)$ (see
Fig.~\ref{fig:Attacks}).
To measure the attack tolerance of networks, we use two
different ways: nodes are deleted either (i) following a fixed decreasing degree
order of the initial network before attacks [Fig.~\ref{fig:Attacks}(c)]
or (ii) following an updated degree order recalculated at every step of single
node removal [Fig.~\ref{fig:Attacks}(d)]. As one can see, the
attacks are more efficient than the errors in destroying the networks.
The attack based on recalculated degree order is the most
efficient, and  the attack based on an initial degree order comes next,
followed by errors on nodes and errors on links. For both errors and
attacks, obviously, AP networks are more fragile than RP networks in
both static and growing cases, as shown in Fig.~\ref{fig:Attacks}. The
GRP network is the most robust, followed by the GAP, the SRP, and then the GRP networks, in this order.
The vulnerability of AP networks can probably be explained by the BCs of AP
networks being larger than those of RP networks. When a link (or node) of high BC is
removed, it is more probable for the largest cluster to rapidly shrink. As shown
in Fig.~\ref{fig:snapshots}, AP networks are more chain-like and have larger BC nodes than RP networks. Therefore, AP networks are more fragile than
RP networks.

\section{Summary and Discussion}
We investigate the structural properties of Achlioptas process (AP) networks by comparing them with
random process (RP) networks for both static and growing cases. The degree distribution and
the local clustering coefficient as functions of degree do not change significantly
for the AP. However, the distributions of the shortest path length and the
betweenness centrality change noticeably for both static and growing networks.
These changes come from the suppression effect of the AP, which prevents the
formation of large clusters, in other words, prohibits inter-community links in the sense of community formation. As a result, the network structure of the AP is more chain-like and more fragile than the network structure of the RP.

\begin{acknowledgments}
This work was supported by a National Research Foundation of Korea (NRF) grant
funded by the Ministry of Science, ICT \& Future Planning No. 2012R1A1A1012150
(SWS) and No. 2014R1A2A2A01004919 (BJK).
\end{acknowledgments}


\begin{references}
\bibitem{note0} In the same sense, $N$ nodes are randomly wired with $L$ links.
\bibitem{note1} In order to study the network structure of the macroscopic-size of the
giant component that exists above the percolation threshold $\delta_c$,
we use the link density $\delta=L/N=1$ (correpsonding to the average degree
 $\langle k \rangle = 2$), which is larger than $\delta_c$ for all four processes considered in this work.
\end{references}
\end{document}